\documentclass[titlepage,12pt]{utarticle}
\usepackage{epsfig}
\usepackage{amsmath,amsfonts}
\usepackage{array}
\usepackage[nosort]{cite}

\numberwithin{equation}{section}

\begin{document}

\preprint{UTTG--02--98\\
{\tt hep-th/9803048}\\
}

\title{Matrix Theory on ALE Spaces and Wrapped Membranes}

\author{David Berenstein and Richard Corrado
        \thanks{Research supported in part by the Robert A.\ Welch
        Foundation and NSF Grant PHY~9511632.}}
\oneaddress{Theory Group, Department of Physics\\
        University of Texas at Austin\\
        Austin TX 78712 USA  \\ {~}\\
        \email{david@zippy.ph.utexas.edu}
        \email{rcorrado@zippy.ph.utexas.edu}
        }

\date{March 5, 1998}

\Abstract{We study the properties of wrapped membranes in matrix
theory on ALE spaces. We show that the only BPS bound states of wrapped
membranes that can form are roots of the $A$-$D$-$E$ group. We
determine a bound on the energy of a bound state and find the correct
dependence on the blow-up parameters and longitudinal momentum
expected from M-Theory. For the $A_{n-1}$ series, we
construct explicit classical solutions for the wrapped membrane bound
states. These states have a very rich structure and have a natural
interpretation in terms of noncommutative geometry.
In the $A_1$ case, we examine the spectrum of excitations
around the wrapped membrane solution and provide an explicit
calculation of their energies. The results agree exactly with supergravity
calculations.}

\maketitle

\renewcommand{\baselinestretch}{1.25} \normalsize

\section{Introduction}

The M(atrix)-Theory~\cite{BFSS:Conjecture} proposal for a
non-perturbative description of M-Theory has been demonstrated to
properly capture M-Theory physics in a variety of settings (see  
\cite{Banks:Matrix-Theory,Bigatti:Review-Matrix,Taylor:Lectures} for
recent reviews). However, several important areas where our
understanding of matrix theory is incomplete still remain. For
example, the fact that matrix theory captures the physics of
linearized supergravity is fairly well
understood~\cite{Kabat:Linearized}. The reasons for its apparent
failure to capture supergravity results in other
situations~\cite{Douglas:Issues,Dine:Multigraviton,Douglas:Hard,%
Becker:Scattering} seem to be subtleties in our understanding
of the discrete light-cone quantization (DLCQ) of M-Theory, namely
that the low-energy description of DLCQ M-Theory is not quite
supergravity~\cite{Banks:Matrix-Theory,Hellerman:Lightlike,%
Bigatti:Review-Matrix}.  

Another area which requires further study is the
compactification of matrix theory, in particular on curved manifolds.
One promising case is that of ``compactification'' on ALE spaces,
which has been conjectured to be described by the theory of D0-brane
partons moving on the ALE space~\cite{Douglas:EnhancedMat,Douglas:Issues,%
Fischler:MatString-K3,Douglas:Strings97}.

In this paper we continue the study of matrix theory on ALE spaces. In
Section~\ref{sec:review}, we give a very brief review of ALE matrix
theory. Our emphasis will be to further examine the characteristics of
the description of membranes which are wrapped around homology
2-cycles as recently described
in~\cite{Diaconescu:Matrix-Mirror,Diaconescu:Fractional-Branes,%
Berenstein:ALE-Matrix}. In
Section~\ref{sec:bound}, we focus on the existence of wrapped membrane
BPS bound states in the wrapped membrane matrix model presented
in~\cite{Berenstein:ALE-Matrix}. We demonstrate that such bound states
must be roots under the $A$-$D$-$E$ group. We then derive a bound on
the energy of such a bound state. For the case of the $A_{n-1}$
series, we explicitly construct the bound states for all roots. In
Section~\ref{sec:noncomm}, we 
briefly discuss how the membrane solution fits into the framework of
noncommutative geometry and thereby satisfies the properties of
spherical membranes~\cite{deWit:QMSupermembrane,Kabat:Spherical}. In
Section~\ref{sec:excitations}, we discuss the spectrum of excitations
of the wrapped membrane and calculate the energies of excitations
around the $A_1$ solution, finding agreement with supergravity expectations.

\section{A Brief Review of ALE Matrix Theory}
\label{sec:review}

A matrix description of M-Theory on an ALE space must possess several
crucial ingredients if it is to be considered both correct and
useful. A candidate formalism for such a description has been provided
by the worldvolume effective theories describing D0-branes moving on
an ALE
space~\cite{Douglas:Quivers,Polchinski:Tensors-K3,Johnson:IIB-ALE}.
Let us briefly review the construction of these models and examine the  
spacetime features of M-Theory that they capture.

\begin{figure}
\centerline{\epsfxsize=8.5cm \epsfbox{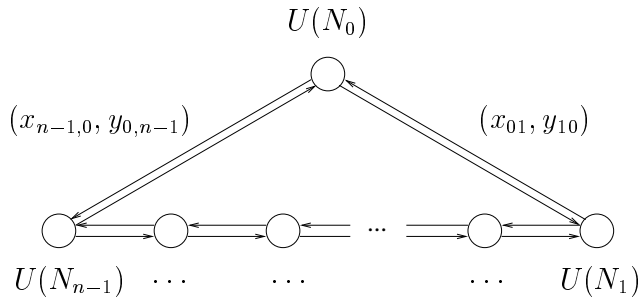}}
\caption{The $A_{n-1}$ quiver diagram, with representative $U(N_i)$
gauge groups at the vertices and hypermultiplets,
$(x_{i,i+1},y_{i+1,i})$, on the edges.} \label{fig:an-diag} 
\end{figure}

The ALE matrix models are given by the dimensional reduction to
quantum mechanics of the six-dimensional gauged supersymmetric sigma
models appearing in the hyperk\"ahler quotient
construction~\cite{Hitchin:Hyperkahler,Kronheimer:ALE} of the ALE
space. The field content is summarized by a quiver diagram, which is
based on the $A$-$D$-$E$ extended Dynkin diagram, such as the
$A_{n-1}$ diagram shown in Figure~\ref{fig:an-diag}. 
For each vertex of the diagram there is a gauge group $U(N_i)$, where
$N_i=N k_i$, with $N$ the number of D0-branes and $k_i$ the Dynkin
label of the vertex, as well as a vector multiplet, $(A_i,a_i)$ (in
terms of $d=4$, $\CN=1$ vector and chiral superfields), in the
$(1,\ldots,\text{ad}(U(N_i)),\ldots,1)$ representation. For each edge
of the diagram there is an associated hypermultiplet,
$(x_{i,i+1},y_{i+1,i})$, in the fundamental--anti-fundamental
representation, $(1,\ldots,1,N_i,\overline{N}_{i+1},1,\ldots,1)$, of
the neighboring gauge groups.   

The Lagrangian is the most general one compatible with
the gauge symmetry and $d=6$, $\CN =1$ SUSY. However, since we are in
$d=1$ dimensions, the question of the mass dimension of the fields and
coupling constants is an issue. We will choose string units, with 
$T_A=1$.  All fields have mass dimension $-1$ and the coupling constant
has dimension $-\frac{1}{2}$. This requires us to scale the
hypermultiplets so that an inverse square of a coupling constant
appears out in front of the terms in the Lagrangian that they appear
in. We will choose this to be the ``average'' coupling constant, defined by 
\begin{equation}
\frac{1}{g^2} = \sum_i \frac{k_i}{g_i^2}. \label{eq:avg-g}
\end{equation}
We also scale the D and F-terms by $2/g^2$. In terms of
$d=4$, $\CN=1$ superfields, our Lagrangian is 
\begin{equation}
\begin{split}
\CL =  \sum_i & \left[  
\left( \frac{1}{16 N_i g_i^2} \int d^2\theta \, W_i^2 + \text{c.c.} \right) 
+ \frac{1}{g_i^2} \int d^4\theta \, \bar{a}_i e^{A_i} a_i e^{-A_i}
\right. \\  
& + \frac{1}{g^2} \int d^4\theta \, 
\left( \bar{x}_{i,i+1} e^{A_i} x_{i,i+1} e^{-A_{i+1}}
+  \bar{y}_{i+1,i} e^{A_{i+1}} y_{i+1,i} e^{-A_i} \right) \\
& + \left( \frac{1}{g^2} \int d^2\theta \, 
\left( y_{i+1,i} a_i x_{i,i+1} - x_{i-1,i} a_i y_{i,i-1} \right) +
\text{c.c.} \right) \\ 
& \left. + \frac{2}{g^2} \int d^4\theta \, d_i A_i 
+ \left( \frac{2}{g^2} \int d^2 \theta \, f_i a_i + \text{c.c.}
\right) \right].
\end{split} \label{eq:lagrangian}
\end{equation}
The coupling constant satisfies
\begin{equation}
\frac{1}{g^2} = T_{D0}. \label{eq:couplingconst} 
\end{equation}
Then it is clear that the $1/g_i^2$ are the masses of the fractional
D0-branes~\cite{Polchinski:Tensors-K3,Douglas:EnhancedMat,%
Diaconescu:Fractional-Branes} associated to each vertex.

The coefficients of the D and F-terms are constrained by the
requirement that a supersymmetric ground state exists. This requires
that 
\begin{equation}
\sum_i k_i f_i = \sum_i k_i d_i =0. \label{eq:vanishing}
\end{equation}
For $N_i=N k_i$, there are zero-energy ground states. Under the
$SU(2)_R$ R-symmetry, the $\vec{\zeta}_i = (f_i,d_i)$ form
triplets. In fact, according to the hyperk\"ahler quotient
construction, the $\vec{\zeta}_i$ are the blow-up parameters for the
ALE space. With the normalization chosen in~\eqref{eq:lagrangian}, the
area of the $i^{th}$ $\BP^1$ is given by $4\pi |\vec{\zeta}_i|^2$.
 
For simplicity, one typically chooses all the gauge
couplings to be the same, but this is not required. As we will see
later, these coupling constants have an interpretation as Wilson lines
for the $A$-$D$-$E$ gauge group in the DLCQ M-Theory description.  

Applying the matrix prescription~\cite{BFSS:Conjecture}, one is lead
to conjecture that, in the limit $N\rightarrow\infty$, $g\rightarrow
\infty$,  the D0-branes described by~\eqref{eq:lagrangian} are the
partons of the infinite-momentum frame description of M-Theory on the
ALE space~\cite{Douglas:EnhancedMat,Douglas:Issues}. On the other hand,
the finite~$N$, finite~$g$ matrix models would be conjectured to
provide a description of the DLCQ of 
M-Theory~\cite{Susskind:AnotherConj} on the ALE
space~\cite{Douglas:Hard}.  

Several pieces of evidence support the M-Theory interpretation of
these ALE matrix models. First, the models contain the geometry of the
ALE space. For $N=1$, we have the standard $U(1)$ gauged
supersymmetric sigma model. After taking the hyperk\"ahler
quotient~\cite{Hitchin:Hyperkahler}, 
we recover the target space $\BR^5\times \CM_{\vec{\zeta}}$. 
Here we use $\CM_{\vec{\zeta}}$ to denote the ALE space parameterized
by the blowup parameters $\vec{\zeta}$.  For larger $N$,
the moduli space of the theory is 
\begin{equation}
( \BR^5\times \CM_{\vec{\zeta}} )^N/S_N,
\end{equation}
so that the ALE geometry is recovered along the flat directions of the
classical ground states of the system.

The models have the equivalent of $\CN=1$ supersymmetry in six
dimensions, or eight supercharges, which is the same amount of
supersymmetry present in light-cone M-Theory on an ALE space. The
gauge group $\prod_i U(N_i)$ contains an overall $U(1)$ factor which is a
linear combination of the $U(1)$s for each $U(N_i)$ factor. The vector
associated to this $U(1)$ is decoupled from the rest of the dynamics. In the
matrix model, these decoupled degrees of freedom are associated to the
center-of-mass motion in the transverse space. The amount of
supersymmetry provides for a 256-fold degeneracy in the continuum
spectrum of the theory, which allows these states to be
identified with the gravity multiplet. The SUSY 
might also allow for the existence of non-renormalization theorems
that could lead to correct results for interactions.

It is crucial that each finite mass BPS object in M-Theory have an
explicit description as some state in the matrix quantum
mechanics. Furthermore, whenever BPS branes come together, or a
2-cycle which has a membrane wrapped around it shrinks to zero-size,
an enhanced gauge symmetry must appear in the quantum dynamics.  

The mechanism by which wrapped membranes and enhanced gauge symmetry
appear in the model was introduced by
Douglas~\cite{Douglas:EnhancedMat} and elaborated upon
in~\cite{Douglas:Issues,Diaconescu:Matrix-Mirror,Douglas:Hard,%
Diaconescu:Fractional-Branes,Berenstein:ALE-Matrix}. 
From the Kronheimer construction~\cite{Kronheimer:ALE}, the vertices
of the extended Dynkin diagram are associated to the homology
2-cycles (which are $\BP^1$s) of the ALE
space. In~\cite{Diaconescu:Matrix-Mirror}, it was proposed that, in
the blow-down, states which correspond to wrapped membranes exist on
the Coulomb branch of the quiver gauge theory. In particular, 
a configuration describing a membrane wrapped on the $i^{th}$ $\BP^1$
would be described by a metastable state for which the hypermultiplets
corresponding to that vertex vanished, $x_{i-1,i}=x_{i,i+1}=\cdots
=0$.  This picture was 
further discussed in~\cite{Diaconescu:Fractional-Branes}, where a
proposal was made for its extension to the blown-up ALE spaces.

A different approach was taken in~\cite{Berenstein:ALE-Matrix}, where
it was proposed that the matrix description of membranes wrapped on
these $\BP^1$s could be obtained by considering the ALE matrix models
described above, but now relaxing the condition that $N_i=N
k_i$. Instead, one should take 
\begin{equation}
N_i= Nk_i +r_i, \label{eq:n-values}
\end{equation}
where $N$ is chosen such that the set of $r_i>0$ are as small as
possible.  The resulting matrix model describes $N$ D0-branes 
propagating on a ALE space with $r_i$ membranes wrapped around the
$i^{th}$ $\BP^1$. It is this latter approach which we will consider in
the rest of the paper. 

If all of the $r_i=n k_i$ for some integer $n$ (we can include the
possibility that $n=0$, in which case $N_i=N k_i$), then the
corresponding homology 2-cycle is trivial, and the configuration of
wrapped membranes can decay into a collection of D0-branes. A single
wrapped membrane corresponds to a simple root of the spacetime gauge
group, while the other roots have been conjectured to be given by
bound states of these simple roots~\cite{Douglas:EnhancedMat}. We will
demonstrate this result explicitly in Section~\ref{sec:bound}.

The objective of the following sections is to give a detailed account
of several features of wrapped membranes in the ALE matrix models. We
give a complete description of what BPS bound states of wrapped membranes
form in the quantum mechanical system and then construct explicit bound
state solutions for the $A_{n-1}$ series. We give a short discussion
of the noncommutative geometry properties of the solutions and exhibit
the spherical membrane properties for the explicit $A_1$
solution. For the $A_1$ ALE space, we examine the spectrum of
excitations of the wrapped membrane, providing an explicit calculation
of their energies. We find that the large $N$ behavior of the excitations
matches supergravity expectations. 

\section{Bound States of Wrapped Membranes}
\label{sec:bound}

We would like to determine the conditions under which a
(non-threshold) BPS bound state of membranes can form. These are ground
states of the interacting part of the theory that, classically,
completely break the gauge symmetry group. Since the argument is
classical, any solution of the F-terms for the hypermultiplets can be
deformed to set the gauge fields equal to zero. In this way, the
symmetry breaking pattern can be chosen to preserve the $SO(5)$
symmetry of the directions transverse to the ALE. 

In order to determine the masses of the gauge fields, the argument can
be extended to the six-dimensional quiver gauge theory. The Higgs
mechanism requires that the extra degrees of freedom that give mass to
a vector multiplet come from ``eating'' a hypermultiplet. The number
of vector multiplets that can be Higgsed can be obtained by counting
the total number of vector multiplet degrees of freedom 
($4 \sum_i N_i^2$) and subtracting the number of hypermultiplet 
($4 \sum_{i,j \,\text{adj.}} N_i N_j$) and decoupled $U(1)$ degrees of
freedom (4). The resulting number must be smaller than zero or otherwise
there remains a flat direction corresponding to a vector multiplet, in
which case the membrane configuration can be separated classically
into two separate bound states.  We therefore need $N_i$ such that
\begin{equation}
0 \geq 
4 \left( \sum_i N_i^2 -\sum_{i,j \, \text{adj.}} N_i N_j - 1 \right)
=  2 \left( \sum_{i,j} \hat{C}_{ij} N_i N_j -2 \right),
\label{eq:eating}
\end{equation} 
where $\hat{C}_{ij}$ is the extended Cartan matrix of the
corresponding $A$-$D$-$E$ group. One immediately recognizes the
quantity $\sum_{i,j} \hat{C}_{ij} N_i N_j$ as the squared norm of the
vector $N_i$ in the lattice of the algebra. Since the
inequality~\eqref{eq:eating} is satisfied whenever 
$\sum_{i,j} \hat{C}_{ij} N_i N_j \leq 2$, we see 
that the state must either be neutral under the Cartan subalgebra (and
is therefore a collection of D0-branes) or it is a root.

It is fairly straightforward, in general, to
determine a bound on the energy of the state. Wrapped membrane
bound states should form massive vector multiplets in seven spacetime
dimensions. Since the fermionic zero-modes corresponding to the
decoupled $U(1)$ center-of-mass motion already give rise to a 16-fold
degeneracy, the ground state of the interacting part of the theory must be 
non-degenerate. This is difficult to prove in general, but we can
still obtain a bound on the energy rather easily. We will then show
that these states exist explicitly in the case of the $A_{n-1}$ ALE spaces.

Let us assume that the F-terms are zero, then, after computing the traces,
the Hamiltonian for the D-terms we obtain from~\eqref{eq:lagrangian} is
\begin{equation}
H_D= \sum_i \left( \frac{N_iD_i^2 }{2g_i^2} - \frac{2N_i d_i D_i}{g^2}
\right). 
\label{eq:dtermham}
\end{equation}
Since the decoupled $U(1)$ is the sum of the $U(1)$s at each vertex, the
corresponding D-term, $D_{\text{dec.}}$, is given by the sum
\begin{equation}
D_{\text{dec.}} = 
\frac{1}{\sum_i \frac{N_i}{g_i^2}} \sum_i \frac{N_iD_i}{g_i^2} .
\label{eq:ddec}
\end{equation}
Therefore, each D-term has an expansion as
\begin{equation}	
D_i = D_{\text{dec.}} + D_{i,\text{int.}}, 
\label{eq:dtermexp}
\end{equation}
where the $D_{i,\text{int.}}$ are orthogonal to each other and
$D_{\text{dec.}}$ and satisfy
\begin{equation}
\sum_i \frac{N_i D_{i,\text{int.}}}{g_i^2}  = 0.
\end{equation}

Using the expansion~\eqref{eq:dtermexp} in~\eqref{eq:dtermham}, we
find that the Hamiltonian for the decoupled $U(1)$ is 
\begin{equation}
H_{\text{dec.}}= D_{\text{dec.}} 
\sum_i N_i \left( \frac{D_{\text{dec.}}}{2g_i^2} - \frac{2d_i}{g^2} \right).
\end{equation}
This is minimized by
\begin{equation}
D_{\text{dec.}} = - \frac{2\sum_i N_i d_i}{g^2 \sum_i \frac{N_i}{g_i^2}}, 
\label{eq:vev-ddec}
\end{equation}
which leads to a bound on the energy
\begin{equation}
E \geq \frac{2\left(\sum_i N_i d_i\right)^2}{g^4\sum_i \frac{N_i}{g_i^2}}. 
\label{eq:bound}
\end{equation}
From this expression for the bound, we see that the
D0-brane charge of the state is proportional to 
$\sum_i \frac{N_i}{g_i^2}$. If we take the $N_i$ as in~\eqref {eq:n-values}
and apply~\eqref {eq:vanishing}, we find that
\begin{equation}
E \geq \frac{2\left(\sum_i r_i d_i\right)^2}{g^4\sum_i \frac{N_i}{g_i^2}}. 
\end{equation}

A similar calculation works when the F-terms are non-zero. Defining
$F_{\text{dec.}}$ in a manner analogous to~\eqref{eq:ddec}, we find
the solution
\begin{equation}
F_{\text{dec.}} = - \frac{2\sum_i N_i f_i}{g^2 \sum_i \frac{N_i}{g_i^2}}.
\label{eq:vev-fdec}
\end{equation}
Putting the results together, we find a bound 
\begin{equation}
E \geq \frac{2}{g^4\sum_i \frac{N_i}{g_i^2}} 
\left( \bigl(\sum_i r_i d_i\bigr)^2 
+ \big|\sum_i r_i f_i \bigr|^2 \right)
= \frac{2\bigl| \sum_i r_i \vec{\zeta}_i\bigr|^2}{g^4 
\sum_i \frac{N_i}{g_i^2}}.  
\label{eq:boundwithf} 
\end{equation}
This is the light-cone energy of an object with mass 
\begin{equation}
m = \frac{2 \bigl| \sum_i r_i \vec{\zeta}_i \bigr|}{g^2}
\label{eq:mem-mass} 
\end{equation}
and longitudinal momentum 
\begin{equation}
p^+ = \sum_i \frac{N_i}{g_i^2}. \label{eq:p-plus}
\end{equation} 
In particular, it is
clear that by tuning the $g_i$, one assigns different amounts of
longitudinal momentum to each simple root. In the DLCQ, the parameters
responsible for this are the Wilson lines of the spacetime gauge group
around the light-like circle. 
			
In the large $N$ limit, we obtain from~\eqref{eq:boundwithf}
\begin{equation}
E \geq \frac{ 2 \bigl| \sum_i r_i \vec{\zeta}_i\bigr|^2}{N g^2} 
\left( 1 - O(1/N) \right).
\end{equation}
We will show explicitly below that there are solutions that satisfy
the bound for any $N$. Moreover, from the DLCQ interpretation
of the coupling constants, we know that at large $N$ the effects of
the Wilson lines should disappear. In particular this implies that the
ratios $g_i/g_j$ become unimportant in the large $N$ limit and
therefore that some of these coupling constants decouple from the
dynamics in the large $N$ limit. This can serve as a very useful
calculational tool, and might serve to produce some
non-renormalization theorems in the large $N$ limit. 

As a check on our results, we can calculate the tension of
the membrane from~\eqref{eq:mem-mass}. For a single membrane state
\begin{equation}
m = \frac{2 \bigl| \vec{\zeta} \bigr|}{g^2} = T^{(2)} A,
\end{equation}
Since the area of the 2-sphere is $A=4\pi | \vec{\zeta} |$,
from~\eqref{eq:couplingconst} 
we recover (in string units) the well known
result~\cite{Green-Hull-Townsend:recursion} 
\begin{equation}
T^{(2)} = \frac{T_{D0}}{2\pi}.
\end{equation}

Now we will construct the bound states for all roots in the $A_{n-1}$
matrix models. First, as the $k_i=1$ for the $A_{n-1}$, the
$r_i=0~\text{or}~1$. For a state to be a root and non-trivial in
homology, all of the $r_i=1$ must be adjacent and there must be at
least one $r_i=0$. 

We will begin with the simplifying assumption that the F-terms are set
to zero. Later, we will describe how to obtain solutions when both
F and D-terms are present. 
In order to solve the F-term equations of the hypermultiplets, we set
all of the $a_i=0$ in the vector multiplets. To solve the F-term
equations of the vector multiplets, we make the ansatz 
\begin{equation}
x_{i,i+1} y_{i+1,i} = y_{i+1,i} x_{i,i+1} =0. \label{eq:nof-ansatz}
\end{equation}  

Now we must minimize the D-terms. The states we are interested in
should be excited states under the decoupled $U(1)$, but should form
supersymmetric bound states of the internal part of the
theory. Therefore, we will set $D_{i,\text{int.}}=0$ and seek a
solution to the equations
\begin{equation}
\bar{x}_{i-1,i} x_{i-1,i} - y_{i,i-1} \bar{y}_{i,i-1} 
- x_{i,i+1} \bar{x}_{i,i+1} + \bar{y}_{i+1,i} y_{i+1,i} 
= \left(  2d_i + \frac{g^2}{g_i^2} D_{\text{dec.}} \right)\, 
\Bid_{N_i}. \label{eq:dterm}
\end{equation}

With our ansatz, the operators $\bar{x}_{i-1,i} x_{i-1,i}$ and
$y_{i,i-1} \bar{y}_{i,i-1}$ commute. As they are self-adjoint, they
can be simultaneously diagonalized, and they moreover multiply to
zero. Therefore they are fully determined as the projections to the
positive and negative spectra of the operator
\begin{equation}
O_i = \bar{x}_{i-1,i} x_{i-1,i} - y_{i,i-1} \bar{y}_{i,i-1}
\end{equation}
Now, $O_i$ and the operator
\begin{equation}
O^\prime_i = x_{i-1,i} \bar{x}_{i-1,i} - \bar{y}_{i,i-1} y_{i,i-1}
\end{equation}
which appears in the D-term at the $(i-1)^{th}$ vertex, are
isospectral, except for an extra zero eigenvalue on one of them
whenever $r_i=r_{i-1}\pm 1$, so that $N_i=N_{i-1}\pm 1$. Imposing an
ordering on the eigenvalues, $O_i=\text{diag}(O_i^1, \ldots,O^{N_i}_i)$,
with $O_i^1 \geq \cdots \geq O^{N_i}_i$, the D-term
equations~\eqref{eq:dterm} become 
\begin{equation} 
O_i^\ell - O_{i+1}^{\prime \ell} 
=  \left( 2d_i + \frac{g^2}{g_i^2} D_{\text{dec.}}   \right)
\label{eq:dtermsbyeigen} 
\end{equation}

Now, if the state is nontrivial, then, without loss of generality, we
can label the first vertex where there is a membrane wrapped as the
$0^{th}$, so that on the quiver diagram we will have the sequence
$\ldots,r_{n-1}=0,r_0=1,r_1=1,\ldots,r_\iota=1,
r_{\iota+1}=0,\ldots$ 
({\em i.e.}, there are $\iota$ 1s and $n-\iota$ 0s on the diagram). Then
$O_0^{N+1}=0$, so that~\eqref{eq:dtermsbyeigen} can be solved to find 
$O_1^{\prime N+1}= -(2d_0 + \frac{g^2}{g_0^2} D_{\text{dec.}})$. But then
$O_1^{N+1}= -(2d_0+ \frac{g^2}{g_0^2} D_{\text{dec.}})$ as well. By
repeated use of~\eqref{eq:dtermsbyeigen}, we can follow these
eigenvalues until we get to the transition $r_\iota=1 \rightarrow
r_{\iota+1}=0$, where we jump to the next non-zero eigenvalue, so that
\begin{equation}
O_{\iota+1}^{N} = O_{\iota+1}^{\prime N+1}
= -\sum_{j=1}^{\iota+1} 
\left( 2 d_{j-1} + \frac{g^2}{g_{j-1}^2} D_{\text{dec.}}  \right).
\end{equation}
By iterating this process around to the $0^{th}$ vertex, we
find 
\begin{equation}
\begin{split}
O_0^N & = - \sum_{j=1}^{n} 
\left( 2d_{j-1} + \frac{g^2}{g_{j-1}^2} D_{\text{dec.}} \right)
= - {D_{\text{dec.}}}      \\  
O_{1}^N &= - 2d_0 
- \left( 1 +\frac{g^2}{g_0^2} \right) D_{\text{dec.}}.
\end{split}
\end{equation} 
By continuing this process of circuiting the diagram, we determine all
of the eigenvalues.  We find the general expression 
\begin{equation}
O^{\ell}_{k} = - \sum_{j=1}^{k} 
\left( 2 d_{j-1} + \frac{g^2}{g_{j-1}^2} D_{\text{dec.}} \right)
- \begin{cases}
(N+1-\ell)  D_{\text{dec.}} & \text{for}~0 \leq k \leq \iota \\
(N-\ell) D_{\text{dec.}} & \text{for}~\iota < k < n. 
\end{cases}
\label{eq:osolution}
\end{equation}
The explicit form for $D_{\text{dec.}}$ is given by the
solution~\eqref{eq:vev-ddec}. For $n=1$, this formula is in agreement
with the $A_1$ solution presented in~\cite{Berenstein:ALE-Matrix}.

Now let us consider the situation when both F and D-terms are
present. Here, in addition to solving the D-terms~\eqref{eq:dterm}, we
must also solve the F-term equations, which are of the form
\begin{equation}
y_{i,i-1} x_{i-1,i} - x_{i,i+1} y_{i+1,i} 
= \left( 2f_i + \frac{g^2}{g_i^2} F_{\text{dec.}} \right)\, \Bid_{N_i}. 
\label{eq:fterm}
\end{equation}
The simplest ansatz for a solution would clearly be one for which the
above method of tracing eigenvalues around the quiver diagram would
work. This will require that the individual terms in~\eqref{eq:dterm}
and~\eqref{eq:fterm} commute amongst themselves, so that they are
simultaneously diagonalizable. We must therefore demand that
\begin{equation}
\begin{split}
\bar{y}_{i,i-1} y_{i,i-1} x_{i-1,i} 
&= x_{i-1,i} y_{i,i-1}\bar{y}_{i,i-1} \\
y_{i,i-1} x_{i-1,i} \bar{x}_{i-1,i} 
&= \bar{x}_{i-1,i} x_{i-1,i} y_{i,i-1}.
\end{split} \label{eq:comm-ansatz}
\end{equation} 

If we then define
\begin{equation}
\begin{split}
P_i &= y_{i,i-1} x_{i-1,i}  \\
P_i^\prime &= x_{i-1,i} y_{i,i-1},
\end{split}
\end{equation}
then it is a straightforward exercise to show that
\begin{equation}
[O_i,P_i] = [P_i,\bar{P}_i]=0.
\end{equation} 
From~\eqref{eq:comm-ansatz}, we see that $x_{i-1,i}$ generates a sort of
``supersymmetry'' for the matrices $\bar{y}_{i,i-1} y_{i,i-1}$ and
$y_{i,i-1}\bar{y}_{i,i-1}$. That is, it acts as an intertwiner between
their eigenspaces, so that given an eigenstate, $|\alpha\rangle$, of
$y_{i,i-1}\bar{y}_{i,i-1}$, the state $x_i|\alpha\rangle$ is an
eigenstate of $\bar{y}_{i,i-1} y_{i,i-1}$ with the same eigenvalue.
Similarly, 
$y_{i,i-1}$ is a supersymmetry for $x_{i-1,i} \bar{x}_{i-1,i}$ and
$\bar{x}_{i-1,i} x_{i-1,i}$, while both $x_{i-1,i}$ and $y_{i,i-1}$
are supersymmetries for $P_i$ and $P_i^\prime$. We can therefore
conclude that, as before, $O_i$ and 
$O_i^\prime$ have the same spectrum, except, perhaps, for a
zero-mode. Similarly, the spectra of $P_i$ and $P_i^\prime$ can differ
only by a zero-mode. 

Solutions may be explicitly obtained by solving the
eigenvalue equations~\eqref{eq:dtermsbyeigen} and
\begin{equation}
P_i^\ell - P_{i+1}^{\prime \ell} 
= \left( 2f_i + \frac{g^2}{g_i^2} F_{\text{dec.}} \right),
\end{equation}
via following the eigenvalues around the quiver diagram, as outlined
above.  Then we find that the $O_i$ are given by~\eqref{eq:osolution},
as before, while the
\begin{equation}
P^{\ell}_{k} = -\sum_{j=1}^{k} 
\left( 2f_{j-1} + \frac{g^2}{g_{j-1}^2} F_{\text{dec.}}\right)
- \begin{cases}
     (N+1-\ell)      F_{\text{dec.}} & \text{for}~0 \leq k \leq \iota \\
     (N-\ell)      F_{\text{dec.}} & \text{for}~\iota < k < n, 
\end{cases}
\label{eq:psolution}
\end{equation}
where $F_{\text{dec.}}$ is given by~\eqref{eq:vev-fdec}.
We note that if the F-terms are in fact absent, we recover our old
ansatz~\eqref{eq:nof-ansatz}. 

Within the ansatz~\eqref{eq:comm-ansatz} the solutions obtained
from~\eqref{eq:osolution} and~\eqref{eq:psolution} are
unique. Explicit expressions for the $x_{i,i+1}$ and $y_{i+1,i}$ can
be obtained by choosing, say, the $x_{i,i+1}$ to have positive real
entries. This can 
always be achieved by a gauge transformation. The solution obtained in
this manner is a classical solution to the equations for a
supersymmetric vacuum (of the internal theory), and therefore there is
a corresponding quantum mechanical state whose wavefunction is
localized near the classical solution. It is clear that all of the
hypermultiplets have a mass gap and it is reasonable to believe that
the solution breaks the gauge group completely. 

\section{Noncommutative Geometry and Spherical Membranes}
\label{sec:noncomm}

It is worthwhile to illustrate the
geometric nature of the membrane solutions we have constructed. We
will show in this section that the natural set of gauge-invariant
coordinates derived from the quiver theory are elevated by these
solutions to noncommuting coordinates\footnote{For a recent discussion
of noncommutative geometry and matrix theory,
see~\cite{Connes:NonComm-Mat} and references therein.}. These
noncommuting coordinates 
will satisfy the constraints imposed by the ALE space geometry to
leading order in the $1/N$ expansion. 

In the case of the $A_{n-1}$, we know that $\BZ_{n}$-invariant
coordinates are given by invariant products of the coordinates of the
$\BC^{n}$ being quotiented. In terms of the quiver theory describing a
single membrane wrapped on the $0^{th}$ $\BP^1$, we can
consider the $N_0\times N_0$ matrices
\begin{equation}
\begin{split}
U &= x_{01}\cdots x_{n-1,0} \\
V &= y_{0,n-1} \dots y_{10} \\
W & = x_{01} y_{10}. 
\end{split} \label{eq:matrix-coords}
\end{equation}
Coordinates $(u,v,w)$ on an ALE space of the same shape will satisfy 
\begin{equation}
uv=P(w), \label{eq:an-poly}
\end{equation}
where $P(w)$ is an $n^{th}$-order polynomial whose coefficients are
determined by the $f_i$. 

Now, when the $f_i=0$ ({\it e.g.}\ on the blow-down), $P(w)=w^n$. Also,
from~\eqref{eq:psolution} we see that
$W=0$. Since~\eqref{eq:osolution} tells us that $O_0$ is
positive-definite (for $\sum_i r_i d_i>0$, $D_{\text{dec.}}$ is
negative-definite), we know that  
$\bar{y}_{01} y_{01} =0$, whence $y_{01}=V=0$. On the other hand,
$O_1$ is negative-definite, so that $x_{01}\bar{x}_{01}=0$ and
$x_{01}=U=0$ as well. Therefore the membrane is indeed localized at
the singularity, as $u=v=w=0$ is the singular locus of~\eqref{eq:an-poly}.

By means of the F-term equations~\eqref{eq:fterm}, one can show that
for large $N$
\begin{equation}
UV = P(W) + O(1/N). 
\end{equation}
Moreover, one can also show that 
\begin{equation}
\begin{split}
[W,U] & = F_{\text{dec.}} U  
=  - \frac{2\sum_i d_i f_i}{N} U + O(1/N^2) \\
[W,V] & = - F_{\text{dec.}} V  
=  \frac{2\sum_i d_i f_i}{N} V  + O(1/N^2).
\end{split}
\end{equation}
This is reminiscent of the angular momentum
commutation relations, and shows that the intrinsic geometry of the
wrapped membranes that we have constructed is noncommutative. The
membranes are spherical, but if we were to probe the membrane locally,
where it approaches the flat membrane 
solution, we would find an effective ``Planck constant'' of order
$1/N$ times the area of the wrapped membrane. This comes as no
surprise when we consider the way membranes were first constructed in
\cite{deWit:QMSupermembrane,BFSS:Conjecture}, as well as the discussion
of spherical membranes in matrix theory by Kabat and
Taylor~\cite{Kabat:Spherical}. 

As an illustration of the noncommutative properties of the solutions,
it is illuminating to check that single membranes have properties
which are analogous to those discussed 
in~\cite{Kabat:Spherical}. In the $A_1$ case, we can set the F-terms
to zero and use~\eqref{eq:osolution} to solve for the hypermultiplets.
Then $X_{01}=Y_{01}=0$, while the other hypermultiplet components are
off-diagonal (up to a $U(N)\times U(N+1)$ gauge transformation)  
\begin{equation} 
\begin{split}
X_{10} &= \sqrt{\frac{4d}{2N+1}} 
\begin{pmatrix} 
\sqrt{N} & 0          & \cdots & \cdots & 0      \\
0        & \sqrt{N-1} & 0      & \cdots & \vdots \\
\vdots   &  0         & \ddots & 0     & \vdots \\ 
0        &  \cdots    & 0      & 1      & 0     
\end{pmatrix} \\
Y_{10} &= \sqrt{\frac{4d}{2N+1}} 
\begin{pmatrix} 
0      & 1       & 0          & \cdots     & 0      \\
\vdots & 0       &  \ddots    & 0         & \vdots \\
\vdots & \cdots       & 0 &      \sqrt{N-1}  & 0 \\ 
0      & \cdots  &  \cdots    & 0          & \sqrt{N}   
\end{pmatrix},
\end{split}
\label{eq:memsolution}
\end{equation}
where we have set $g_0=g_1$ for convenience. As coordinates on the ALE
space are gauge-invariant combinations of the quiver theory
hypermultiplet components, we may consider the quadratic combinations 
\begin{equation}
\begin{split}
J_0  &= \frac{1}{8\sqrt{d}} 
\bigl( X_{10} \bar{X}_{10}  - Y_{10} \bar{Y}_{10} \bigr) \\
&= \frac{\sqrt{d}}{2N+1} 
\begin{pmatrix}
N-1 & 0 & \cdots & \cdots & 0 \\
0 & N-3 & 0 & \cdots & \vdots \\
\vdots & 0 & \ddots & 0 & \vdots  \\
\vdots    & \cdots & 0       &  -(N-3) & 0         \\
0       &  \cdots  &\cdots      &    0    & -(N-1)  
\end{pmatrix} 
\end{split} \label{eq:su2gen0}
\end{equation}
\begin{equation}
J_+ = \frac{1}{4\sqrt{d}}  Y_{10} \bar{X}_{10}
= \frac{2\sqrt{d}}{2N+1} 
\begin{pmatrix}
0          & \sqrt{N-1}        & 0    &  \cdots     & 0 \\
\vdots & 0             & \sqrt{2(N-2)}    &  0    & \vdots \\
\vdots    &  \cdots & 0         &\ddots & 0 \\
\vdots     & \cdots     & \cdots &    0   & \sqrt{N-1} \\
0          & \cdots        &  \cdots        & \cdots & 0 \\ 
\end{pmatrix} \label{eq:su2gen+}
\end{equation}
\begin{equation}
J_- = \frac{1}{4\sqrt{d}} X_{10} \bar{Y}_{10} 
= \frac{2\sqrt{d}}{2N+1} 
\begin{pmatrix}
0          & \cdots        & \cdots    &  \cdots     & 0 \\
\sqrt{N-1} & 0             & \cdots    &  \cdots     & \vdots \\
0          & \sqrt{2(N-2)} & 0         &\cdots & \vdots \\
\vdots     & 0             & \ddots &    0   & \vdots \\
0          & \cdots        &  0        & \sqrt{N-1} & 0 \\ 
\end{pmatrix}  . \label{eq:su2gen-}
\end{equation}
These matrices are proportional to the generators of
the $\textbf{N}$ of $SU(2)$, so they commute according to the
$SU(2)$ algebra 
\begin{equation}
\begin{split}
& [ J_+,J_-]  = 2 \left( \frac{2\sqrt{d}}{2N+1} \right) J_0 \\
& [ J_\pm,J_0]  = \mp \left( \frac{2\sqrt{d}}{2N+1} \right) J_\pm
\end{split}
\end{equation}
and the sum over their squares is
\begin{equation}
\sum_i J_i^2= J_0^2 + \frac{1}{2} \{ J_+, J_- \} 
= \frac{4(N^2-1)d}{(2N+1)^2} \Bid_N = r^2 \Bid_N.
\label{eq:sumoversq} 
\end{equation}
Evidently, \eqref{eq:su2gen0}--\eqref{eq:su2gen-} are the redundant set
of coordinates 
which parameterize the surface of the membrane. We note that the set
of operators $\tilde{J}_+= \bar{X}_{10} Y_{10} /4\sqrt{d},\ldots$
also form a set of membrane coordinates, this time in
the~$\mathbf{N+1}$ representation. That there is a pair of good
membrane coordinates can be useful in calculations. 

As in~\cite{Kabat:Spherical}, we can ask if there is a regime in which
these membranes approach the flat membrane solution. At large $N$,
\begin{equation}
[ J_+,J_-] = \frac{d}{N} 
\begin{pmatrix} \Bid & 0 \\ 0 & -\Bid \end{pmatrix} + O(1/N^2),
\end{equation}
which, when restricted to the upper-left quadrant, resembles the
commutation relations of the coordinates of the flat membrane.
In the semi-classical correspondence between Poisson brackets and
commutators of~\cite{deWit:QMSupermembrane,BFSS:Conjecture}, we find
the ``Planck constant''
\begin{equation}
\hbar = \frac{d}{N},
\end{equation} 
which, as promised above, is proportional to the area of the membrane.

In the large $N$ limit, from~\eqref{eq:sumoversq}, we also find that
$r$ is the radius of the membrane,   
\begin{equation}
r = \sqrt{d} \left( 1 - \frac{1}{2N} \right).
\label{eq:radius}
\end{equation} 
Our choice of normalization in~\eqref{eq:su2gen0}--\eqref{eq:su2gen-}
was made specifically 
to obtain this leading-order behavior. These results agree nicely with
those of~\cite{Kabat:Spherical}, once the different $N$ dependence of
the membrane longitudinal momentum here is taken into account. We note
that the corrections to the radius~\eqref{eq:radius} are stronger than
those found for the membranes in~\cite{Kabat:Spherical}, being of
$O(1/N)$ as compared to $O(1/N^2)$.  

\section{The Spectrum of Excitations of the Wrapped Membrane}
\label{sec:excitations}

We will now concentrate our efforts on finding the spectrum of
excitations around the membrane solution for the simplest possible
case, namely the $A_1$ singularity. We will first examine a toy model,
in order to establish a framework for understanding the problem. Then
we utilize the rotational invariance of the $A_1$ solution to
determine the allowed representations for perturbations around the
membrane. We then calculate the mass spectrum for the P-wave
excitations and, finally, for all modes.

We note that the perturbative approach we take to the calculation of
the spectrum of fluctuations is well-suited only for membranes which
are large as compared to the 11-dimensional Planck length. It is only
in this limit that we may reliably expect that the fluctuations have
energies small compared to the mass of the membrane and are thereby
long-lived.  

We want to describe the linearized spectrum of excitations of a large
membrane wrapped around a sphere. Consider a toy model consisting of a
massless free scalar field on the space $S^2\times \BR$, where $\BR$
is time and the sphere is of radius~$r$. It is a straightforward
exercise to show that the normal modes are the spherical harmonics,
with masses given by
\begin{equation}
M_\ell = \frac{\sqrt{\ell(\ell+1)}}{r} \label{eq:toy}
\end{equation}
for $\ell=0,1,$ etc. This toy model reproduces the linearized degrees
of freedom for a BPS membrane wrapped on a 2-sphere, if we give the
action maximal supersymmetry. Instead of a single scalar, we can
consider $U(1)$ SYM with maximal supersymmetry as our starting point. 
This is the natural effective worldvolume theory of a D2-brane (see,
{\it e.g.}, \cite{Polchinski:NotesDbranes,Polchinski:TASIDbrane}) and,
in particular, is valid for the large radius limit we are considering.

As we are considering a free $U(1)$ theory, there is no need to
quantize the magnetic flux through the sphere. This quantum number is
associated to the compactness of the 11th dimension. In order to get
an explicitly $O(6)$-invariant Lagrangian, as one would expect from
M-Theory, one can dualize the vector into a scalar.

Since half of the supersymmetries are broken in the ALE space, we must also
explicitly break half of the supersymmetry of our toy model. This can
be achieved by twisting one complex scalar. In particular, we identify
one of the scalars as a section of the normal bundle of the
2-sphere sitting in the ALE space. The resulting theory is
supersymmetric if the fermions are sections of the bundle
$\CO(-1)\oplus \CO(1)$. There are six bosonic zero modes, which
correspond to the motion of the membrane in the transverse $\BR^6$.
The corresponding eight fermionic zero modes generate the required
16-fold degeneracy of the ground state.  For the massive modes, there
are eight bosonic and eight fermionic modes for each spherical
harmonic. This gives us the 
information we require in order to interpret the calculation we will
now make in the matrix model.  

For the $A_1$ singularity there is an $SU(2)$
rotational invariance which aids in the solution. Namely, since there
is a pair of hypermultiplets connecting  
the same two vertices, the action (and so too the D and F-terms) is
invariant under the $SU(2)_r$ transformation
\begin{equation}
\begin{pmatrix} x_{01} \\  y_{01} \end{pmatrix} \rightarrow 
\begin{pmatrix} x_{01}^\prime \\  y_{01}^\prime \end{pmatrix} 
= R \begin{pmatrix} x_{01} \\  y_{01} \end{pmatrix} ,~~~
\begin{pmatrix} x_{10} \\  y_{10} \end{pmatrix} \rightarrow 
\begin{pmatrix} x_{10}^\prime \\  y_{10}^\prime \end{pmatrix} 
= R \begin{pmatrix} x_{10} \\  y_{10} \end{pmatrix},
\label{eq:rotation}
\end{equation}
for $R\in SU(2)_r$, which mixes the hypermultiplets.
Clearly there is a $U(N)\times U(N+1)$ gauge
transformation,
\begin{equation} 
\begin{pmatrix} U & 0 \\  0 & U \end{pmatrix} R 
\begin{pmatrix} X_{10} \\ Y_{10} \end{pmatrix} 
V 
= \begin{pmatrix} X_{10} \\ Y_{10} \end{pmatrix},
\end{equation} 
where $U\in U(N)$ and $V\in U(N+1)$, that undoes the action
of~\eqref{eq:rotation} on the membrane ground
state~\eqref{eq:memsolution}. In other words, 
the solution~\eqref{eq:memsolution} is invariant under the product
$U\otimes R\otimes V$, which is a consequence of the fact that the
ground state is spherically symmetric. Obviously $U$ and $V$ must
themselves correspond to (inverse) elements of $SU(2)_r$. 

Now the action of $U\otimes R\otimes V$ on any perturbations
around~\eqref{eq:memsolution} should be faithful. Since $R$ acts as
the $\textbf{2}$ of $SU(2)_r$ on 
$\Bigl(\begin{smallmatrix} \delta x_{10} \\ \delta y_{10} 
\end{smallmatrix}\Bigr)$ 
and $U$ and $V$ act as the $\textbf{N}$ and $\textbf{N+1}$,
respectively, $U\otimes R \otimes V$ must act as the representation 
\begin{equation}
\textbf{2}  \otimes \textbf{N}\otimes(\textbf{N+1})  
= \textbf{1} \oplus \textbf{3} \oplus \textbf{3} \oplus \textbf{5}
\oplus \textbf{5} \oplus \cdots \oplus \textbf{2N--1} \oplus
\textbf{2N--1}\oplus\textbf{2N+1}.   \label{eq:irreps-hyper}
\end{equation}
On the other hand, under this gauge transformation,
the vector multiplets transform in the 
\begin{equation}
\begin{split}
\textbf{N}\otimes\textbf{N} &= \textbf{1} \oplus \textbf{3} \oplus
\cdots \oplus \textbf{2N--1} \\
\textbf{(N+1)}\otimes(\textbf{N+1}) &= \textbf{1} \oplus \textbf{3} \oplus
\cdots \oplus \textbf{2N+1}  
\end{split} \label{eq:irreps-vectors}
\end{equation}
representations. 

One immediately notices that the dimension of the representation of
the vectors~\eqref{eq:irreps-vectors} is greater by a singlet than that
of the hypermultiplets in~\eqref{eq:irreps-hyper}. This singlet is
simply the decoupled $U(1)$. The absence of a singlet for the
decoupled $U(1)$ in the hypermultiplet representation protects against
the possibility of Higgsing away this element of the gauge
symmetry. Another important observation 
about~\eqref{eq:irreps-hyper} and~\eqref{eq:irreps-vectors} is that
only even spherical harmonics will appear in the multipole
expansion of the perturbations. This is a considerable simplification,
and we will be able to compute the full spectrum of excitations of the
wrapped membrane. 

As a start, let us now calculate the mass spectrum for the P-wave
excitations, {\em i.e.}, those in the~$\textbf{3}$ of~$SU(2)_r$. The
calculation will be illustrative of the techniques we will use to
compute the spectrum for generic $j$. 

Since the masses are generated via the Higgs mechanism, it will be
sufficient to calculate  the masses of the vector multiplets, since
these are related to those 
of the hypers in an obvious way. Furthermore, we can exploit the
$SO(5)$ invariance of the scalars in the vector multiplet, so that we
have $a_{0i} = a^\alpha_0\lambda^{(N+1)}_\alpha$, 
$a_{1i} = a^\alpha_1\lambda^{(N)}_\alpha$, for each~$i$, where
the $\lambda^{(N)}_\alpha$ generate the~$\textbf{N}$
of~$SU(2)_r$. Finally, we note from the~$SU(2)_r$ invariance that we
can take $a^\alpha_0=a_0$, $a^\alpha_1=a_1$, for each $\alpha$. We
then can compute the mass of $a_0$ and $a_1$ for the simplest case,
namely the Cartan generator, $\lambda_0$. 
 
The quadratic part of the Lagrangian for $a_0$ and $a_1$ is
\begin{equation}
\begin{split}
\CL  =&
\frac{1}{2g_0^2} \, \text{Tr} \bigl( \lambda^{(N+1)}_0 \bigr)^2 \dot{a}_0^2 
+ \frac{1}{2g_1^2} \, \text{Tr}\bigl( \lambda^{(N)}_0 \bigr)^2 \dot{a}_1^2 \\
& - \frac{1}{g^2} \left( \text{Tr}\left[ \bigl( \lambda^{(N+1)}_0 \bigr)^2 
(\bar{X}_{10}X_{10} + \bar{Y}_{10} Y_{10}) \right] a_0^2
+ \text{Tr}\left[ \bigl( \lambda^{(N)}_0 \bigr)^2 
(X_{10}\bar{X}_{10} + Y_{10}\bar{Y}_{10}) \right] a_1^2 \right. \\
& \phantom{+\frac{1}{g^2}}~ \left. + 2 \, \text{Tr} \left[ 
\lambda^{(N)}_0 X_{10}  \lambda^{(N+1)}_0 \bar{X}_{10}
+ \lambda^{(N)}_0 Y_{10}  \lambda^{(N+1)}_0 \bar{Y}_{10} \right] a_0
a_1 \right).
\end{split} \label{eq:quadlag}
\end{equation}
From~\eqref{eq:memsolution}, if we generalize to arbitrary coupling
constants at each vertex, we have
\begin{equation}
\begin{split}
X_{10}\bar{X}_{10} + Y_{10}\bar{Y}_{10} 
& = (N+1) |D_{\text{dec.}}| \, \Bid_N \\
\bar{X}_{10}X_{10} + \bar{Y}_{10} Y_{10}
& = N |D_{\text{dec.}}| \, \Bid_{N+1} \\
X_{10}\bar{X}_{10} - Y_{10}\bar{Y}_{10} 
& = |D_{\text{dec.}}| \, \lambda^{(N)}_0 \\
\bar{X}_{10}X_{10} - \bar{Y}_{10} Y_{10}
& =  |D_{\text{dec.}}| \, \lambda^{(N+1)}_0.
\end{split} \label{eq:sums}
\end{equation}
To simplify the cross-terms
in~\eqref{eq:quadlag}, we use the ``commutation'' relations 
\begin{equation}
\begin{split}
\lambda^{(N)}_0 X_{10} - X_{10} \lambda^{(N+1)}_0 + X_{10} &= 0 \\
\lambda^{(N)}_0 Y_{10} - Y_{10} \lambda^{(N+1)}_0 - Y_{10} &= 0. 
\end{split} \label{eq:comm}
\end{equation}
These relations encode, in part, the invariance of the membrane
solution~\eqref{eq:memsolution} under the rotation group of the
sphere~\eqref{eq:rotation}. 
One may also readily calculate 
\begin{equation}
\text{Tr} \bigl( \lambda^{(N)}_0 \bigr)^2 = \frac{(N-1)N(N+1)}{3}.
\label{eq:tracesq}
\end{equation}

Using these relations in~\eqref{eq:quadlag} yields
\begin{equation}
\CL = \frac{1}{2} \bigl( \dot{\tilde{a}_0}^2 + \dot{\tilde{a}_1}^2 \bigr)
-\frac{|D_{\text{dec.}}|}{g^2}
\left[ N g_0^2 \tilde{a}_0^2 
+ 2 \sqrt{(N-1)(N+2)} \, g_0 g_1 \tilde{a}_0 \tilde{a}_1
+ (N+1) g_1^2 \tilde{a}_1^2 \right], 
\label{eq:normlag}
\end{equation}
where we have defined the normalized variables
\begin{equation}
\begin{split}
\tilde{a}_0 &= 
\sqrt{ \text{Tr} \bigl( \lambda^{(N+1)}_0 \bigr)^2 } \, \frac{a_0}{g_0} \\
\tilde{a}_1 &= 
\sqrt{ \text{Tr} \bigl( \lambda^{(N)}_0 \bigr)^2 } \, \frac{a_1}{g_1}.
\end{split} 
\end{equation}
The eigenvalues of the mass matrix are the energies for the oscillator modes
\begin{equation}
\begin{split}
\omega_\pm^2 &= \left[ \left( Ng_0^2 + (N+1) g_1^2 \right) 
\pm \sqrt{ \left( Ng_0^2 + (N+1) g_1^2 \right)^2 - 8 g_0^2 g_1^2}
\right]  \frac{ |D_{\text{dec.}}| }{g^2}\\
& \sim \frac{2 g_0^2 g_1^2 d}{g^4} 
\left[ 1 \pm \left( 1 - \frac{4g^4}{g_0^2 g_1^2} \, \frac{1}{N^2} 
+ \cdots \right) \right].
\end{split} \label{eq:energies-3}
\end{equation}

Therefore, in the large $N$ limit, we find two modes, one whose energy
is independent of $N$ at leading order (and therefore decouples) and
another whose energy is 
\begin{equation}
\omega_- = \frac{2\sqrt{2d}  }{N    }
\end{equation}
and survives as the membrane excitation as $N\rightarrow \infty$. 
Notice that the energy of this mode only depends on $d$,
providing some evidence for the hypothesis made in
Section~\ref{sec:bound} that the ratio $g_0/g_1$ 
decouples in the large $N$ limit . The mass of this mode, $\delta m$,
can be calculated from 
\begin{equation}
\omega_- = \frac{(m+\delta m)^2-m^2}{2p^+} 
\sim \frac{m \, \delta m}{p^+}, 
\end{equation}
where $m$ and $p^+$ are given by~\eqref{eq:mem-mass}
and~\eqref{eq:p-plus}, respectively. In this way, we find
\begin{equation}
\delta m = \sqrt{\frac{2}{d}} = \frac{\sqrt{2}}{r},
\end{equation}
where $r$ is the radius of the membrane. This result agrees precisely 
with our toy model result~\eqref{eq:toy} for the P-wave, $\ell=1$.
                                                 
Now, for generic excitations appearing in the $\mathbf{2j+1}$
representation under the decomposition~\eqref{eq:irreps-hyper}, we may
again exploit symmetry arguments. In general, these reduce the problem
to computing the masses of the scalars in the vector multiplets, which
may be taken to be of the form
\begin{equation}
\begin{split}
a_{0i} &= a_0 \bigl( \lambda^{(N+1)}_+ \bigr)^j 
+ \bar{a}_0 \bigl( \lambda^{(N+1)}_- \bigr)^j \\
a_{1i} &= a_1 \bigl( \lambda^{(N)}_+ \bigr)^j 
+ \bar{a}_1 \bigl( \lambda^{(N)}_- \bigr)^j,
\end{split}
\end{equation}
for each $i=1,\ldots,5$. As a motivation for why we have chosen this
particular form, we note that these are the highest weight states for
their respective $SU(2)_r$ representations. This stems from the fact
that, as the representations are irreducible, all of the operators
acting in a given representation are constructed from polynomials in
the generators.   

After dropping the terms which vanish after
taking traces, the quadratic Lagrangian for these modes is
\begin{equation}
\begin{split}
\CL = & 
2 \, \text{Tr} \left[ 
\bigl( \lambda^{(N+1)}_+ \bigr)^j \bigl( \lambda^{(N+1)}_- \bigr)^j 
\right]  \, \left( \frac{1}{g_0^2} |\dot{a}_0|^2 
-  \frac{N |D_{\text{dec.}}|}{g^2} |a_0|^2 \right) \\
& + 2 \, \text{Tr} \left[ 
\bigl( \lambda^{(N)}_+ \bigr)^j \bigl( \lambda^{(N)}_- \bigr)^j 
\right]  \, \left( \frac{1}{g_1^2} |\dot{a}_1|^2 
- \frac{(N+1) |D_{\text{dec.}}|}{g^2} |a_1|^2 \right) \\
& - \frac{2}{g^2} a_0 \bar{a}_1 \left[ 
\bigl( \lambda^{(N)}_+ \bigr)^j X_{10} 
\bigl( \lambda^{(N+1)}_- \bigr)^j \bar{X}_{10}
+ \bigl( \lambda^{(N)}_+ \bigr)^j Y_{10} 
\bigl( \lambda^{(N+1)}_- \bigr)^j \bar{Y}_{10} \right] \\
&  -\frac{2}{g^2} \bar{a}_0 a_1 \left[ 
\bigl( \lambda^{(N)}_- \bigr)^j X_{10} 
\bigl( \lambda^{(N+1)}_+ \bigr)^j \bar{X}_{10}
+ \bigl( \lambda^{(N)}_- \bigr)^j Y_{10} 
\bigl( \lambda^{(N+1)}_+ \bigr)^j \bar{Y}_{10} \right] .
\end{split} \label{eq:lag-generic}
\end{equation} 
The cross-terms here can be simplified through the use of the
``commutation'' relations 
\begin{equation}
\begin{split}
X_{10} \lambda^{(N+1)}_+ - \lambda^{(N)}_+ X_{10} - Y_{10} &= 0 \\ 
X_{10} \lambda^{(N+1)}_- - \lambda^{(N)}_- X_{10} &= 0 \\
Y_{10} \lambda^{(N+1)}_+ - \lambda^{(N)}_+ Y_{10} &= 0 \\
Y_{10} \lambda^{(N+1)}_- - \lambda^{(N)}_- Y_{10} - X_{10}  &= 0. 
\end{split} \label{eq:comm-pm}
\end{equation}
As in the case of the relations~\eqref{eq:comm}, these encode the
rotational invariance of the membrane solution. By repetitive
application of~\eqref{eq:comm-pm}, one may establish the useful relationship
\begin{equation}
(N-j) \, \text{Tr} \left[ 
\bigl( \lambda^{(N+1)}_+ \bigr)^j \bigl( \lambda^{(N+1)}_- \bigr)^j
\right]
= (N+j+1) \, \text{Tr} \left[ 
\bigl( \lambda^{(N)}_+ \bigr)^j \bigl( \lambda^{(N)}_- \bigr)^j
\right].
\end{equation}

In terms of the normalized variables
\begin{equation}
\begin{split}
\tilde{a}_0 &= 
\sqrt{ 2 \,\text{Tr} \left[ 
\bigl( \lambda^{(N+1)}_+ \bigr)^j \bigl( \lambda^{(N+1)}_- \bigr)^j 
\right] } \, \frac{a_0}{g_0} \\
\tilde{a}_1 &= \sqrt{ 2 \, \text{Tr} \left[ 
\bigl( \lambda^{(N)}_+ \bigr)^j \bigl( \lambda^{(N)}_- \bigr)^j 
\right] }\, \frac{a_1}{g_1},
\end{split} 
\end{equation}
the Lagrangian~\eqref{eq:lag-generic} becomes
\begin{equation}
\begin{split}
\CL = & |\dot{\tilde{a}_0}|^2 + |\dot{\tilde{a}_1}|^2  
-\frac{|D_{\text{dec.}}|}{g^2} \left( N g_0^2 |\tilde{a}_0|^2 
+ (N+1) g_1^2 |\tilde{a}_1|^2 \right) \\
& - \frac{|D_{\text{dec.}}|}{g^2} \sqrt{(N-j)(N+j+1)} \, g_0 g_1 
\left( \tilde{a}_0 \bar{\tilde{a}_1} 
+ \bar{\tilde{a}_0} \tilde{a}_1 \right),
\end{split} \label{eq:norm-genlag}
\end{equation}
From this, we find the energies of the modes as the eigenvalues of
the mass matrix 
\begin{equation}
\begin{split}
\omega_\pm^2 &= \left[ \left( N g_0^2 + (N+1) g_1^2 \right) 
\pm \sqrt{ \left( N g_0^2 + (N+1) g_1^2 \right)^2 - 4j(j+1) g_0^2 g_1^2}
\right]  \frac{|D_{\text{dec.}}|}{g^2} \\
& \sim \frac{2g_0^2 g_1^2 d}{g^4} 
\left[ 1 \pm \left( 1 - \frac{2j(j+1)g^4}{g_0^2 g_1^2} \, \frac{1}{N^2} 
+ \cdots \right) \right].
\end{split} 
\end{equation}
We note that, in the case $j=1$, this reduces properly
to~\eqref{eq:energies-3}.  

As in the case of the P-waves above, we can consider the behavior
at large $N$. There are again two modes, one which decouples and
another with energy
\begin{equation}
\omega_- = \frac{\sqrt{2j(j+1)d}}{N}
\end{equation}
and mass
\begin{equation}
\delta m =  \sqrt{\frac{j(j+1)}{d}} = \frac{\sqrt{j(j+1)}}{r},
\end{equation}
which agrees exactly with the toy model result~\eqref{eq:toy}. In the
limit of large radius, the effective field theory on the membrane and
the perturbation expansion in the matrix model agree.

\section{Conclusions}

We have given quite a bit of evidence that the matrix model proposed
in~\cite{Douglas:Quivers , Berenstein:ALE-Matrix} captures many of the
essential features of M-Theory membranes which are wrapped around the
homology 2-cycles of an ALE space. In~\cite{Berenstein:ALE-Matrix}, we
gave an explicit derivation of the solution describing the wrapped
membrane in the $A_1$ case. We found that its energy, as well as the
leading order membrane-antimembrane potential, had the properties
necessary for its interpretation as a wrapped membrane. 

In Section~\ref{sec:bound}, we gave a counting argument that set the
correspondence between BPS bound states of membranes and roots of the
$A$-$D$-$E$ group. We derived a general bound on the light-cone energy
of a membrane bound state and found that it had the necessary
dependence on the blow-up parameters and longitudinal momentum. We
also gave an explicit solution for all membrane bound states for the
$A_{n-1}$ series. The 16-fold degeneracy of these solutions is
consistent with the requirement that they form BPS vector multiplets
in seven-dimensional spacetime physics.  

We were able to obtain all of the solutions that we expected from the
M-Theory interpretation of the model, but we have not been able to
show that we have given a unique solution. The existence of other
solutions would probably lead to a greater than 16-fold degeneracy
that would be hard to reconcile with the expected seven-dimensional
physics.   

The solutions we found exhibit a very rich
structure, and fit naturally into a noncommutative geometry framework.
In section~\ref{sec:noncomm}, we showed that there exist gauge-invariant
coordinates for the membrane in the $A_1$ model that satisfy the
relations appropriate to spherical membranes~\cite{Kabat:Spherical}.
  
We then characterized the representations of excitations around the $A_1$
membrane in Section~\ref{sec:excitations}, finding that, as expected,
only even spherical harmonics contribute in the multipole
expansion. We then calculated the energy spectrum of the excitations
and found the exact agreement with the expected result from the
consideration of a large membrane, where we could trust both a
calculation in a toy model, as well as in matrix model perturbation theory.
The end result is very promising. The perturbation analysis gives the
correct results in the large radius limit, and some coupling constants
seem to decouple in the large $N$ limit. If this decoupling is a
generic property of the large $N$ dynamics, this opens up the possibility of
proving non-renormalization theorems.

Despite our success in solving for the membrane bound state
solutions for the $A_{n-1}$ series, we have not been able to
generalize them to the $D$ and $E$ series. This is basically due to
the fact that our method of following the eigenvalues around the
quiver diagram, while perfect for closed quivers, does not work for
the $D$ and $E$ open quivers. By inspection of a few cases where $N$
is taken very small, one may also see that not all of 
the matrices in these cases can be diagonalized simultaneously. This
is bound to complicate finding membrane solutions. 

However, one should obtain similar results for the membrane properties
in these cases. For example, the association of the BPS bound states with
roots of the $A$-$D$-$E$ algebra and the bound on the energy of a
bound state discussed in Section~\ref{sec:bound} are both completely
general, so they apply equally well to the $D$ and $E$ series. Also,
at large $N$, one expects that the couplings $g_i/g_j$ 
should still decouple from the low energy degrees of freedom (whose
energies scale as~$1/N$), as the states that feel these couplings
should have energies of order~1. Of course, even given a solution, in
these cases the mass matrix is more complicated and should prove
difficult to diagonalize. 

There are still more tests that the matrix models for the ALE spaces should 
be put to. For example, it is particularly interesting to reproduce
the Coulomb  
and velocity-dependent potentials between membranes and
gravitons~\cite{Aharony:MemDyn,Lifschytz:MemInt,Kabat:Linearized,%
Kabat:Spherical}. 
These and related matters are currently under investigation.  

\section*{Acknowledgements}

We would like to thank Philip Candelas, Dan Kabat, and especially
Willy Fischler for enlightening discussions about these and related
issues. We thank Jacques Distler for his collaboration at an early
stage of this work and many important conversations.

\renewcommand{\baselinestretch}{1.0} \normalsize


\bibliography{strings,m-theory,susy}

\providecommand{\href}[2]{#2}\begingroup\raggedright\begin{thebibliography}{10}

\bibitem{BFSS:Conjecture}
T.~Banks, W.~Fischler, S.~H. Shenker, and L.~Susskind, ``{M} {T}heory as a
  matrix model: {A} conjecture,'' {\em Phys. Rev.} {\bf D55} (1997) 5112--5128,
  \href{http://xxx.lanl.gov/abs/hep-th/9610043}{{\tt hep-th/9610043}}.

\bibitem{Banks:Matrix-Theory}
T.~Banks, ``Matrix theory,'' \href{http://xxx.lanl.gov/abs/hep-th/9710231}{{\tt
  hep-th/9710231}}.

\bibitem{Bigatti:Review-Matrix}
D.~Bigatti and L.~Susskind, ``Review of matrix theory,''
  \href{http://xxx.lanl.gov/abs/hep-th/9712072}{{\tt hep-th/9712072}}.

\bibitem{Taylor:Lectures}
W.~Taylor, ``Lectures on {D}-branes, gauge theory and {M}(atrices),''
  \href{http://xxx.lanl.gov/abs/hep-th/9801182}{{\tt hep-th/9801182}}.

\bibitem{Kabat:Linearized}
D.~Kabat and W.~Taylor, ``Linearized supergravity from matrix theory,''
  \href{http://xxx.lanl.gov/abs/hep-th/9712185}{{\tt hep-th/9712185}}.

\bibitem{Douglas:Issues}
M.~R. Douglas, H.~Ooguri, and S.~H. Shenker, ``Issues in {M}(atrix) theory
  compactification,'' {\em Phys. Lett.} {\bf B402} (1997) 36--42,
  \href{http://xxx.lanl.gov/abs/hep-th/9702203}{{\tt hep-th/9702203}}.

\bibitem{Dine:Multigraviton}
M.~Dine and A.~Rajaraman, ``Multigraviton scattering in the matrix model,''
  \href{http://xxx.lanl.gov/abs/hep-th/9710174}{{\tt hep-th/9710174}}.

\bibitem{Douglas:Hard}
M.~R. Douglas and H.~Ooguri, ``Why matrix theory is hard,''
  \href{http://xxx.lanl.gov/abs/hep-th/9710178}{{\tt hep-th/9710178}}.

\bibitem{Becker:Scattering}
K.~Becker and M.~Becker, ``On graviton scattering amplitudes in {M} theory,''
  \href{http://xxx.lanl.gov/abs/hep-th/9712238}{{\tt hep-th/9712238}}.

\bibitem{Hellerman:Lightlike}
S.~Hellerman and J.~Polchinski, ``Compactification in the lightlike limit,''
  \href{http://xxx.lanl.gov/abs/hep-th/9711037}{{\tt hep-th/9711037}}.

\bibitem{Douglas:EnhancedMat}
M.~R. Douglas, ``Enhanced gauge symmetry in {M}(atrix) theory,'' {\em JHEP
  Elec. J.} {\bf 7} (1997), no.~4,
  \href{http://xxx.lanl.gov/abs/hep-th/9612126}{{\tt hep-th/9612126}}.

\bibitem{Fischler:MatString-K3}
W.~Fischler and A.~Rajaraman, ``M(atrix) string theory on {K3},'' {\em Phys.
  Lett.} {\bf B411} (1997) 53--58,
  \href{http://xxx.lanl.gov/abs/hep-th/9704123}{{\tt hep-th/9704123}}.

\bibitem{Douglas:Strings97}
M.~R. Douglas, ``D branes and matrix theory in curved space,'' in {\em Strings
  '97 {\textup{(Amsterdam, June 16-21, 1997)}}}.
\newblock \href{http://xxx.lanl.gov/abs/hep-th/9707228}{{\tt hep-th/9707228}}.

\bibitem{Diaconescu:Matrix-Mirror}
D.-E. Diaconescu and J.~Gomis, ``Duality in matrix theory and three-dimensional
  mirror symmetry,'' \href{http://xxx.lanl.gov/abs/hep-th/9707019}{{\tt
  hep-th/9707019}}.

\bibitem{Diaconescu:Fractional-Branes}
D.-E. Diaconescu, M.~R. Douglas, and J.~Gomis, ``Fractional branes and wrapped
  branes,'' \href{http://xxx.lanl.gov/abs/hep-th/9712230}{{\tt
  hep-th/9712230}}.

\bibitem{Berenstein:ALE-Matrix}
D.~Berenstein, R.~Corrado, and J.~Distler, ``Aspects of {ALE} matrix models and
  twisted matrix strings,'' \href{http://xxx.lanl.gov/abs/hep-th/9712049}{{\tt
  hep-th/9712049}}.

\bibitem{deWit:QMSupermembrane}
B.~de~Wit, J.~Hoppe, and H.~Nicolai, ``On the quantum mechanics of
  supermembranes,'' {\em Nucl. Phys.} {\bf B305} (1988) 545.

\bibitem{Kabat:Spherical}
D.~Kabat and W.~Taylor, ``Spherical membranes in matrix theory,'' {\em Adv.
  Theor. Math. Phys.} {\bf 2} (1998), no.~1,
  \href{http://xxx.lanl.gov/abs/hep-th/9711078}{{\tt hep-th/9711078}}.

\bibitem{Douglas:Quivers}
M.~R. Douglas and G.~Moore, ``D-branes, quivers, and {ALE} instantons,''
  \href{http://xxx.lanl.gov/abs/hep-th/9603167}{{\tt hep-th/9603167}}.

\bibitem{Polchinski:Tensors-K3}
J.~Polchinski, ``Tensors from {K3} orientifolds,'' {\em Phys. Rev.} {\bf D55}
  (1997) 6423--6428, \href{http://xxx.lanl.gov/abs/hep-th/9606165}{{\tt
  hep-th/9606165}}.

\bibitem{Johnson:IIB-ALE}
C.~V. Johnson and R.~C. Myers, ``Aspects of type {IIB} theory on {ALE}
  spaces,'' {\em Phys. Rev.} {\bf D55} (1997) 6382--6393,
  \href{http://xxx.lanl.gov/abs/hep-th/9610140}{{\tt hep-th/9610140}}.

\bibitem{Hitchin:Hyperkahler}
N.~J. Hitchin, A.~Karlhede, U.~Lindstrom, and M.~Rocek, ``Hyperk{\"a}hler
  metrics and supersymmetry,'' {\em Commun. Math. Phys.} {\bf 108} (1987) 535.

\bibitem{Kronheimer:ALE}
P.~B. Kronheimer, ``The construction of {ALE} spaces as hyper-{K\"a}hler
  quotients,'' {\em J. Diff. Geom.} {\bf 19} (1989) 665--683.

\bibitem{Susskind:AnotherConj}
L.~Susskind, ``Another conjecture about {M}(atrix) theory,''
  \href{http://xxx.lanl.gov/abs/hep-th/9704080}{{\tt hep-th/9704080}}.

\bibitem{Green-Hull-Townsend:recursion}
M.~B. Green, C.~M. Hull, and P.~K. Townsend, ``D-brane {W}ess-{Z}umino actions,
  {T}-duality and the cosmological constant,'' {\em Phys. Lett.} {\bf B382}
  (1996) 65--72, \href{http://xxx.lanl.gov/abs/hep-th/9604119}{{\tt
  hep-th/9604119}}.

\bibitem{Connes:NonComm-Mat}
A.~Connes, M.~R. Douglas, and A.~Schwarz, ``Noncommutative geometry and matrix
  theory: Compactification on tori,''
  \href{http://xxx.lanl.gov/abs/hep-th/9711162}{{\tt hep-th/9711162}}.

\bibitem{Polchinski:NotesDbranes}
J.~Polchinski, S.~Chaudhuri, and C.~V. Johnson, ``Notes on {D}-branes,''
  \href{http://xxx.lanl.gov/abs/hep-th/9602052}{{\tt hep-th/9602052}}.

\bibitem{Polchinski:TASIDbrane}
J.~Polchinski, ``T{ASI} {L}ectures on {D}-branes,''
  \href{http://xxx.lanl.gov/abs/hep-th/9611050}{{\tt hep-th/9611050}}.

\bibitem{Aharony:MemDyn}
O.~Aharony and M.~Berkooz, ``Membrane dynamics in {M}(atrix) theory,'' {\em
  Nucl. Phys.} {\bf B491} (1997) 184--200,
  \href{http://xxx.lanl.gov/abs/hep-th/9611215}{{\tt hep-th/9611215}}.

\bibitem{Lifschytz:MemInt}
G.~Lifschytz and S.~D. Mathur, ``Supersymmetry and membrane interactions in
  {M}(atrix) theory,'' {\em Nucl. Phys.} {\bf B507} (1997) 621--644,
  \href{http://xxx.lanl.gov/abs/hep-th/9612087}{{\tt hep-th/9612087}}.

\end{thebibliography}\endgroup
\bibliographystyle{utphys}

\end{document}